\documentclass[]{interact}
\usepackage{xcolor}
\usepackage{epstopdf}
\usepackage[caption=false]{subfig}
\usepackage[longnamesfirst,sort]{natbib}%
\usepackage[utf8]{inputenc}
\usepackage{appendix}
\usepackage{pdfpages}
\usepackage[utf8]{inputenc}
\usepackage[english]{babel}
\usepackage{graphicx}
\usepackage{geometry}
\usepackage{amsmath} 
\usepackage{bm}
\usepackage{flafter}
\usepackage{float}
\usepackage{amssymb}
\usepackage{graphicx}
\usepackage{setspace}
\usepackage{array}
\usepackage{multicol}
\usepackage{placeins}
\bibpunct[, ]{(}{)}{;}{a}{,}{,}

\theoremstyle{plain}

\begin{document}
\title{Penalized Likelihood Methods for Modeling Count Data}

\author{
\name{Minh Thu Bui\textsuperscript{a}, Cornelis J. Potgieter\textsuperscript{a,b}, and Akihito Kamata\textsuperscript{c}}
\affil{\textsuperscript{a}Texas Christian University; \textsuperscript{b}University of Johannesburg; \textsuperscript{c}Southern Methodist University}
}

\maketitle

\begin{abstract}
The paper considers parameter estimation in count data models using penalized likelihood methods. The motivating data consists of multiple independent count variables with a moderate sample size per variable. The data were collected during the assessment of oral reading fluency (ORF) in school-aged children. A sample of fourth-grade students were given one of ten available passages to read with these differing in length and difficulty. The observed number of words read incorrectly (WRI) is used to measure ORF. Three models are considered for WRI scores, namely the binomial, the zero-inflated binomial, and the beta-binomial. We aim to efficiently estimate passage difficulty, a quantity expressed as a function of the underlying model parameters. Two types of penalty functions are considered for penalized likelihood with respective goals of shrinking parameter estimates closer to zero or closer to one another. A simulation study evaluates the efficacy of the shrinkage estimates using Mean Square Error (MSE) as metric. Big reductions in MSE relative to unpenalized maximum likelihood are observed. The paper concludes with an analysis of the motivating ORF data.
\end{abstract}

\section{Introduction}

The definition of Oral Reading Fluency (ORF) is ``the oral translation of text with speed and accuracy,'' see for example \cite{fuchs2001oral} and \cite{shinn1992curriculum}. Reading fluency is a skill developed during childhood that  is needed to understand the meaning of texts and literary pieces. There is a strong correlation between reading fluency and reading comprehension, see \cite{allington1983fluency}, \cite{johns1983informal}, \cite{samuels1988decoding}, and \cite{schreiber1991understanding}. According to \cite{disalle2017impact}, once a student has identified a word and read it correctly, their focus generally shifts from word recognition (attempting to recognize the word) to comprehension (making meaning of the word). This leads to overall understanding of the text. These authors have claimed that incompetent ORF levels are the cause of up to 90\% of reading fluency issues. If a child does not read fluently, their ability to read comprehensively is also hindered and they will have trouble in grasping the meaning of texts. Thus, ORF is a method of evaluating whether a child is at their appropriate reading level compared to their peers and assists in identify at-risk students with poor reading skills. 

In this paper, we analyze ORF data collected from a sample of $508$ fourth-grade students. Each child was given one of ten available passage to read and the number of words read incorrectly (WRI) was recorded. This resulted in around 50 WRI measurements per passage. Reading sessions were recorded so that observer error in counting the number of words read correctly and incorrectly could be eliminated. The WRI scores were obtained from these recorded sessions and are assumed free of measurement error. Strong readers tend to have low WRI scores and weak readers tend to have high WRI scores. However, as the passages are not all equal in difficulty, it is important to be cautious in directly using WRI scores obtained from different passages to measure overall ORF levels in a classroom setting. 

Our work is motivated by noting that, to the best of our knowledge, ORF assessment in practice neither makes any adjustments to account for variations in passage difficulty  nor quantifies the differences in passage difficulty. Instead, in implementation a student is given one minute to read as many words as possible in a 250 word passage, after which an assessor calculates their words correct per minute (WCPM) score by subtracting the number of words read incorrectly from the total number of words read. This WCPM score does not make adjustments for passage difficulty and is currently still the most prevalent measure used to assess ORF, see \cite{miura2007literature}, \cite{fuchs2001oral}, and \cite{hasbrouck2006oral}.

The statistical novelty of this work stems from the use of penalized maximum likelihood to estimate parameters in a count data setting where the counts are naturally bounded (below by $0$ and above by passage length). Penalty functions are used to ``encourage'' estimated passage-specific parameters to be close to one another and/or close to zero. This particular implementation of parameter shrinkage is motivated by the structural properties of the data. Firstly, the passages in an ORF assessment differ with respect to vocabulary used and how sentences are constructed. It follows that the passages naturally vary in difficulty, although they are designed to be comparable. Secondly, passages are designed to not be overly challenging for proficient readers, meaning that it is fairly common to have WRI scores of $0$. Finally, passage-specific sample sizes are small relative to the number of passages.

There is, of course, a rich literature on parameter shrinkage in various statistical models. One of the definitive examples in the multivariable setting is the James-Stein estimator of the mean, see \cite{stein1956inadmissibility}. This estimator is often described as ``borrowing'' information between variables to obtain a more efficient estimator. Other applications of shrinkage include \cite{pandey1985bayes} and \cite{jani1991class} who considered univariate Bayes-type shrinkage in, respectively, a Weibull distribution and an exponential distribution. In the bivariate setting, shrinkage was used to estimate probabilities of the form $P(Y<X)$ for underlying exponential distributions, see \cite{baklizi2003shrinkage}.

One of the most frequently encountered applications of shrinkage is in regression models with a large number of predictor variables. The lasso, developed by \cite{tibshirani1996regression}, is one such technique which revolutionized parameter estimation in generalized linear models (GLMs). The lasso shrinks regression parameters towards zero using an $L_1$ penalty, resulting in predictors being “dropped” from the model by setting the corresponding coefficients equal to zero. The lasso was predated by ridge regression which uses an $L_2$ penalty, see \cite{hoerl1970ridge}. This approach results in some regression coefficients being very close to zero, but does not eliminate potential predictor variables from the model altogether. Other examples of shrinkage applied to GLMs include \cite{maansson2013developing} and \cite{qasim2020new} who developed Liu-type estimators for, respectively, a zero-inflated negative binomial regression model and a Poisson regression model. Shrinkage estimation of fixed effects in a random-effects zero-inflated negative binomial model was considered by \cite{zandi2021using}. The monographs by \cite{gruber2017improving} and \cite{hastie2019statistical} are very good resources for further exploration of shrinkage in regression models.

We would be remiss to not highlight the similarity of penalty-based frequentist estimation methods to Bayesian methods with appropriately selected prior functions. For example, \cite{efron1973stein} show how the James-Stein mean estimator belongs to a larger class of empirical Bayes estimators. Similarly, as a parallel to lasso regression, \cite{park2008bayesian} define a Bayesian lasso for sparse regression estimation. For an overview of some of the recent developments in Bayesian regularization using hierarchical models, see \cite{polson2019bayesian}.

In this paper, measures of passage-specific difficulty are of primary interest. The measure of difficulty considered here is $p=\mathrm{E}[\mathrm{WRI}/N]$ with $N$ the passage length. That is, define the proportion of words read incorrectly in a passage as a measure of difficulty. The required expected value can be expressed as a function of the underlying count data model parameters, meaning their estimation is of central importance. Parameter shrinkage applied to count data models has received limited attention in the literature. In the univariate case of estimating a binomial success probability, \cite{lemmer1981note} considered three different estimators of $p$, while \cite{lemmer1981ordinary} proposed estimators of the type $w \hat{p} + (1-w)p_0$ where $p_0$ is an \textit{a priori} guess. However, neither of these papers consider likelihood-based methods nor provide guidance on selecting the amount of shrinkage.

Our literature review brought a few papers to our attention that are similar in spirit, but consider parameter estimation through shrinkage problem from fundamentally different perspectives. In the frequentist paradigm, \cite{hansen2016efficient} considers three shrinking approaches -- restricted maximum likelihood, an efficient minimum distance approach, and a projection approach -- for estimating model parameters. The work of Hansen requires the specification of a shrinkage direction, which is similar to the selection of a penalty function. In the Bayesian paradigm, \cite{agresti2005bayesian} consider hierarchical models for estimating multinomial success probabilities and \cite{datta2016bayesian} consider estimating the intensity parameter of quasi-sparse Poisson count data. The scarcity of relevant literature highlights the opportunities available to further explore shrinkage estimation methods.

The remainder of this paper proceeds as follows. In Section 2, the penalized likelihood approach is more fully developed, emphasizing the binomial distribution for clarity of exposition. In Section 3, V-fold cross-validation is presented as a data-driven approach for selecting the shrinkage level. Section 4 presents results from extensive simulation studies and the motivating data are analyzed in Section 5.

\section{Shrinkage through Penalized Likelihood Methods}\label{PenLik}

\subsection{Shrinkage through Penalized Likelihood Estimation}
	
Consider a collection of random variables $\bm{X}=\{X_{ij}\}$, $j=1,\ldots,n_i$, $i=1,\ldots,I$, with the $X_{ij} \sim F(\cdot|\bm{\theta}_i)$ mutually independent. Here, $F(\cdot|\bm{\theta}_i)$ denotes a distribution function with $p$-dimensional parameter $\bm{\theta}_i \in \bm{\Theta}\subset \mathbb{R}^{p}$. Let $\bm{\Theta}^{I}=\bm{\Theta}\times \cdots \times \bm{\Theta}$ denote the parameter space associated with the collection of parameters $\bm{\theta}=(\bm\theta_1,\ldots,\bm{\theta}_I)$. Also let $\mathcal{\ell}(\bm{\theta}|\bm{X})$ denote the log-likelihood of the data $\bm{X}$ and let $\bm{\mathcal{S}}_0\subseteq \bm\Theta^I$ denote a specified subset of the parameter space that is of interest. Finally, for $\mathbf{s},\mathbf{t}\in \mathbb{R}^{p\times I}$, let $\tilde{h}(\mathbf{s},\mathbf{t})$ be a norm. We then define
$h(\bm\theta|\bm{\mathcal{S}}_0) = \inf_{\bm{t}\in \bm{\mathcal{S}}_0} \tilde{h}(\bm{\theta},\bm{t})$. That is, $h(\bm\theta|\bm{\mathcal{S}}_0)$ is the shortest distance between a point $\bm{\theta}$ and the space $\mathcal{S}_0$ as measured by the norm $h$. Note that whenever $h(\bm\theta_1|\bm{\mathcal{S}}_0)<h(\bm\theta_2|\bm{\mathcal{S}}_0)$, the point $\bm\theta_1$ is closer to the region $\bm{\mathcal{S}}_0$ than the point $\bm\theta_2$. 

In this context, parameter shrinkage is said to be any estimation method that balances adherence to the data-generating model as measured by $\ell(\bm{\theta}|\bm{X})$ and the closeness of any estimator to $\mathcal{S}_0$ as measured by $h(\bm{\theta}|\bm{\mathcal{S}}_0)$. One such approach is penalized maximum likelihood. Adopting the convention that $\mathrm{Pen}(\bm{\theta})=h(\bm{\theta}|\bm{\mathcal{S}}_0)$ denote the penalty function, the penalized likelihood estimator $\tilde{\bm{\theta}}$ is found by minimizing
\begin{equation}
    D(\bm{\theta}) = -\ell(\bm{\theta}|\bm{x}) + \lambda \mathrm{Pen}(\bm{\theta}) \label{eq:D(theta)}
\end{equation}
with $\lambda>0$ a specified constant. The two component functions of $D(\bm{\theta})$ often exist in some kind of tension; minimizing $-\ell(\bm{\theta}|\bm{x})$ gives the maximum likelihood estimator (MLE), while $\mathrm{Pen}(\bm{\theta})$ attains a minimum for any $\bm{\theta}$ in $\bm{\mathcal{S}}_0$ where the desired parameter constraint is fully satisfied. The tension can be ascribed to the MLE not necessarily being close to the subset of interest $\bm{\mathcal{S}}_0$. The magnitude of $\lambda$ determines the balance between these at times competing interest. 

Calculation of the penalized likelihood estimator $\tilde{\bm{\theta}}$ requires the specification of a generating model, a penalty function, and a value for the parameter $\lambda$. Throughout this paper, generating models closely related to the binomial distribution are considered. All models considered naturally accommodate counts restricted to the set $\{0,1,\ldots,N\}$. The remainder of Section 2 will consider some possible choices of the penalty function while assuming $\lambda$ is known, with the choice of $\lambda$ discussed in Section 3. Note that when it comes to the selection of a penalty function, it will often be the case that the subject-matter expert presents the statistician with a non-mathematical description of $\mathcal{S}_0$. There may be multiple ways of constructing a set $\mathcal{S}_0$ and a penalty function $\mathrm{Pen}(\bm{\theta})$ that satisfies the description. Therefore, the penalty functions considered in this paper should not be considered an exhaustive enumeration of the possibilities. Rather, these are intended to illustrate the many ways in which shrinking can be implemented.

\subsection{Shrinkage to Zero in Binomial Models}
\begin{figure}[H]
	\centering
	\begin{minipage}{0.5\textwidth}
		\centering
		\includegraphics[trim={5cm 2.5cm 5cm 18cm},clip,width=1\textwidth]{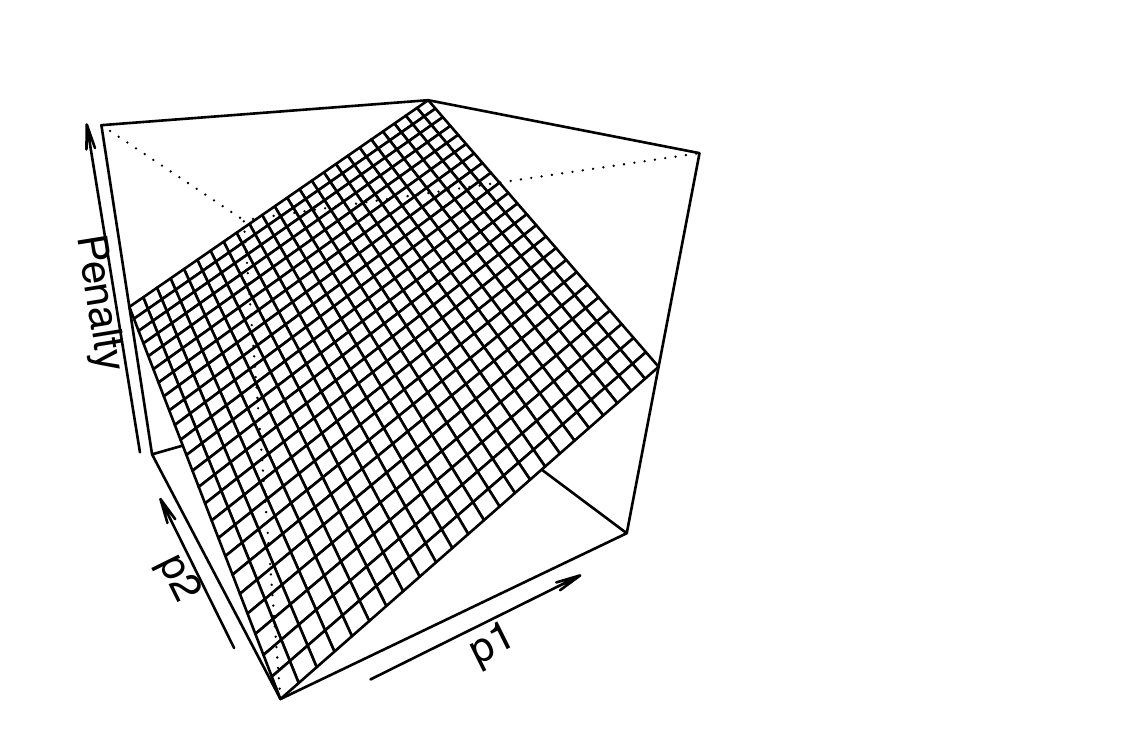} 
	\end{minipage}\hfill
	\begin{minipage}{0.5\textwidth}
		\centering
		\includegraphics[trim={5cm 2.5cm 5cm 18cm},clip,width=1\textwidth]{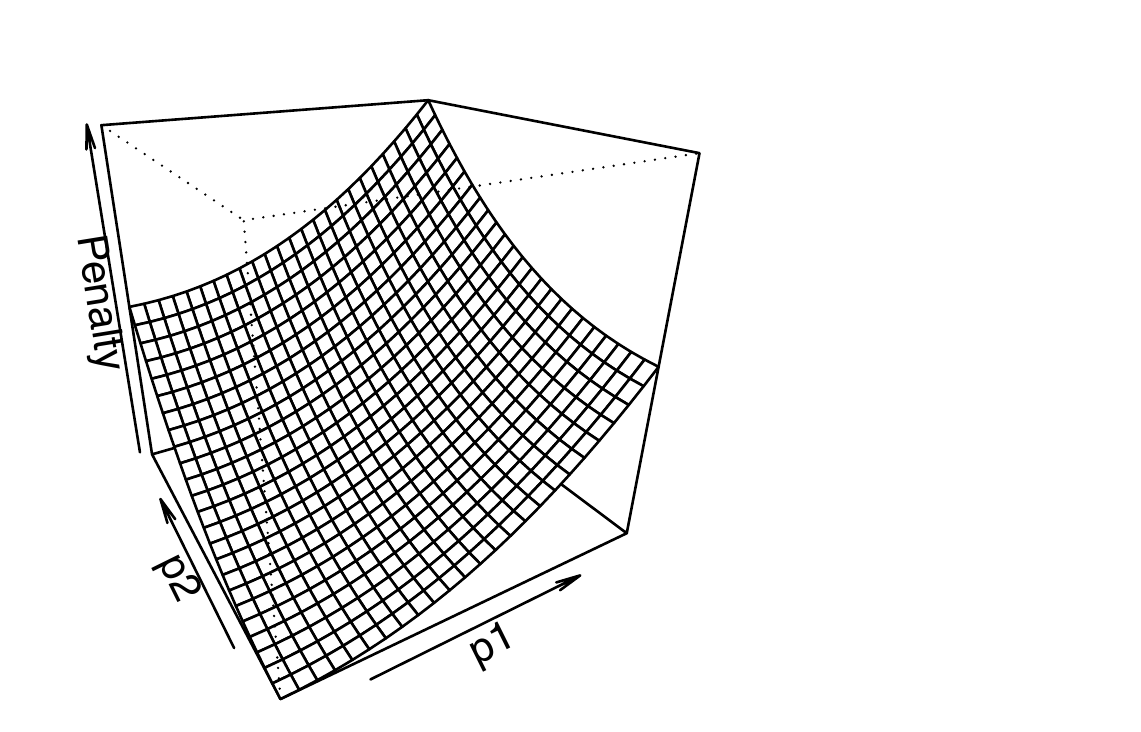} 
	\end{minipage}
\caption{$L_1$ norm (left) and $L_2$ norm (right) penalty functions for $J=2$ binomial success probabilities.}
\label{fig:pen_illustration1}
\end{figure}

Let $x_{i}$, $i=1,\ldots,I$ denote observed realizations of independent random variables $X_i  \sim \mathrm{Bin}(N_i,p_i)$, $i=1,\ldots,I$. Assume that the number of binomial trials $N_i$ are known and that estimation of the success probabilities $p_i$, $i=1,\ldots,I$, is of interest. The log-likelihood is given by
\[\ell(\bm{p}|\bm{x}) = \sum_{i = 1}^{I}\log \binom{N_i}{x_i} + \sum_{i = 1}^{I} x_i \log(p_i) + \sum_{i = 1}^{I} (N_i - x_i)\log(1- p_i).\]
Now, consider the hypothetical scenario where the subject-matter expert has expressed that the success probabilities should all be ``small.'' In the context of the WRI data, this is equivalent to expecting that only a small proportion of words will be read incorrectly by a reader at grade-level. This is consistent with setting $\bm{\mathcal{S}}_0 = (0,\ldots,0)$. There are numerous penalty functions that can assess the closeness of a potential parameter value $\bm{p}=(p_1,\ldots,p_I)$ to $\bm{\mathcal{S}}_0$. For example, both the $L_1$ and $L_2$ norms
\begin{equation}
    \mathrm{Pen}_1(\bm{p}) = \sum_{i=1}^{I} p_i \quad \mathrm{and}\quad \mathrm{Pen}_2(\bm{p}) = \sum_{i=1}^{I} p_i^2, \label{eq:L1_and_L2_pen}
\end{equation}
are candidates worth considering. In the context of binomial success probabilities, both of these functions are bounded, having $\sup_{\bm{p}} \mathrm{Pen}_1(\bm{p}) = \sup_{\bm{p}} \mathrm{Pen}_2(\bm{p})  = I.$ Figure \ref{fig:pen_illustration1} visualizes these penalties for the case $I=2$. The axes $p_1$ and $p_2$ range from $0$ to $1$ in the direction of the arrows. The value of the penalty function itself is omitted from the plot as the magnitude is only informative up to a constant of proportionality. This emphasizes that the goal here (and with other penalty functions graphs that follow) is only to illustrate the shape of these functions.

Note that as $\mathrm{Pen}_2(\bm{p}) \leq \mathrm{Pen}_1(\bm{p})$ for all $\bm{p}\in [0,1]^{I}$, the $L_1$ norm will more aggressively shrink success probabilities to $0$ than the $L_2$ norm. Due to the resemblance of the $L_1$ norm to the commonly-used lasso penalty in regression, it should be pointed out that its application here will not result in shrinkage estimators exactly equal to $0$. In fact, the penalized negative log-likelihood function $D_1(\bm{p}) = -\ell(\bm{p}|\bm{x}) + \lambda \mathrm{Pen}_1(\bm{p})$
has unique solution
\[\tilde{p}_i = \frac{1}{2}\left(\frac{\lambda_i+1}{\lambda_i}\right)\left[1-\left(1-\frac{4\lambda_i\hat{p}_i}{(\lambda_i+1)^2}\right)^{1/2}\right],\ i=1,\ldots,I\]
where $\lambda_i = \lambda/N_i$ and $\hat{p}_i = x_i/N_i$ is the unpenalized MLE. While it is not necessarily intuitive from the form of the penalized estimator, it can easily be verified that $0<\tilde{p}_i <\hat{p}_i$ for all $\lambda>0$. The solution to the $L_2$ penalty function is also easy to compute, but no general closed-form expression is possible as it requires solving a cubic polynomial.

\begin{figure}[H]
	\centering
	\begin{minipage}{0.48\textwidth}
		\centering
		\includegraphics[trim={5cm 2.5cm 5cm 18cm},clip,width=1\textwidth]{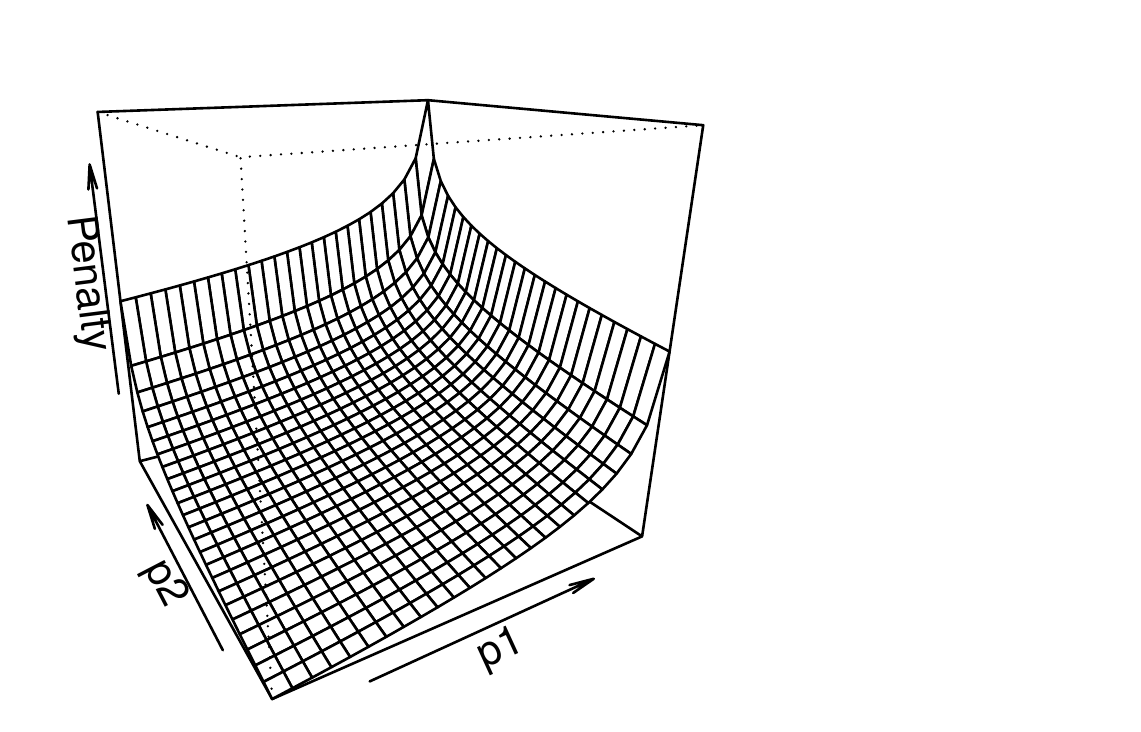} 
	\end{minipage}\hfill
	\begin{minipage}{0.48\textwidth}
		\centering
		\includegraphics[trim={5cm 2.5cm 5cm 18cm},clip,width=1\textwidth]{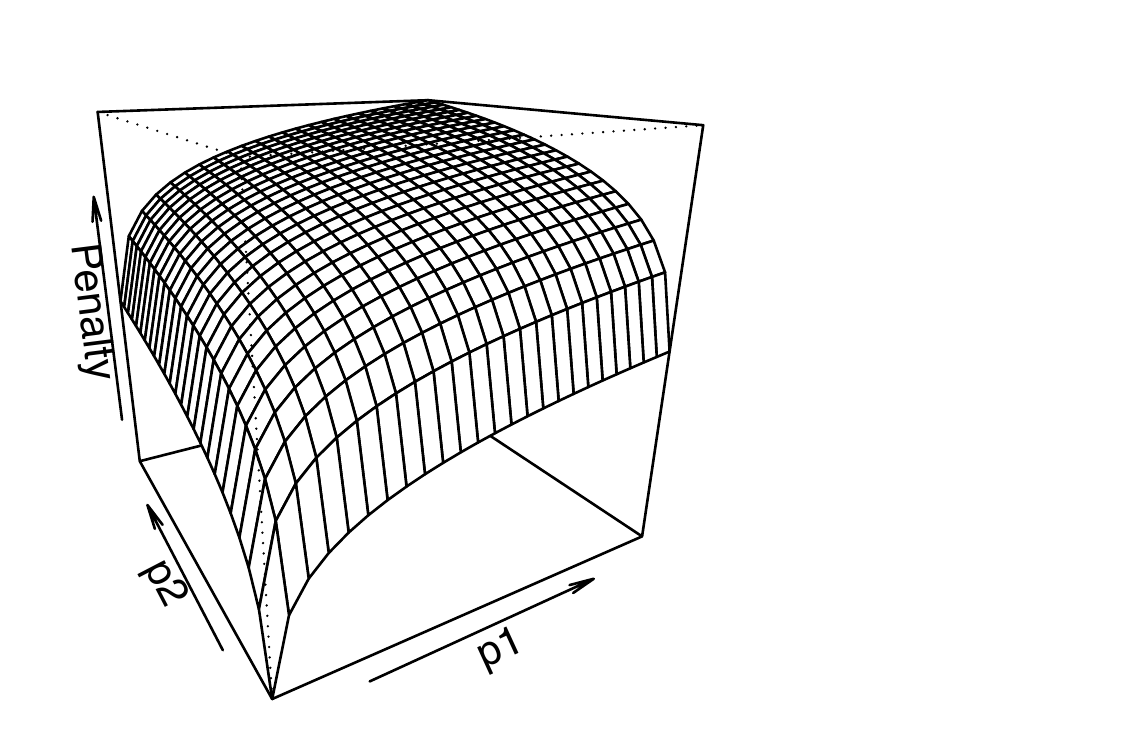} 
	\end{minipage}
	\caption{Penalties $\mathrm{Pen}_3$ (left) and $\mathrm{Pen}_4$ (right) penalty functions, respectively unbounded from above and below, for $I=2$ binomial success probabilities.}
	\label{fig:pen_illustration2}
\end{figure}

The bounded nature of $\mathrm{Pen}_1$ and $\mathrm{Pen}_2$ in \eqref{eq:L1_and_L2_pen} may not appeal to some. One choice of an unbounded penalty is
\[\mathrm{Pen}_{3}(\bm{p}) = - \sum_i \log (1-p_i). \]
This penalty has a lower bound of $0$, but has no upper bound. For an illustration when $I=2$, see Figure \ref{fig:pen_illustration2}. The solution to the corresponding penalized likelihood problem is
\[\tilde{p}_i = \frac{N_i}{N_i+\lambda}\hat{p}_i,\ i=1,\ldots,I.\]
None of the penalties considered so far have the lasso-like property of shrinking parameters to $0$ for a finite value of $\lambda$. However, it is possible to find a penalty that achieves this. Consider
\[\mathrm{Pen}_{4}(\bm{p}) = \sum_i \log p_i \]
also illustrated in Figure \ref{fig:pen_illustration2} for $I=2$. This penalty function is bounded above, but has no lower bound as the individual $p_i$'s approach $0$. In fact, this penalty function is \textit{not} associated with a norm as defined in Section 2.1, putting it somewhat outside the framework in which our estimation problem has been formulated. The latter point notwithstanding, the corresponding penalized likelihood estimator is
\[\tilde{p}_i = \left\{ 
\begin{array}{ll}
	\dfrac{N_i}{N_i-\lambda}\hat{p}_i - \dfrac{\lambda}{N_i-\lambda} & \lambda \leq x_i \\ 
	0 & \lambda > x_i
\end{array}%
\right. \]
for $i=1,\ldots,I$. Perhaps this penalty can appropriately be described as ``greedy'' in the sense that it has the potential to dominate the data and result in a shrinkage estimator equal to $0$ even when there are observed successes suggesting otherwise.

\begin{figure}[H]
	\centering
		\includegraphics[trim={1cm 1.5cm 1cm 18cm},clip,width=0.8\textwidth]{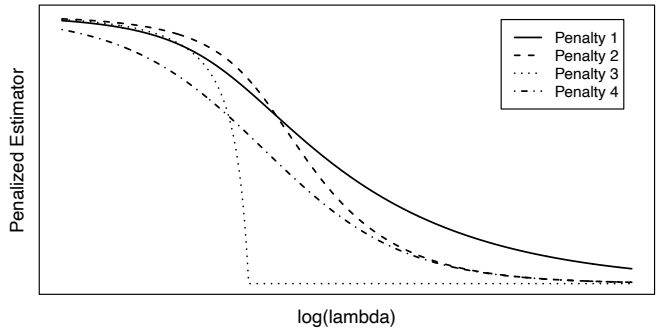} 
	\caption{Schematic representation of four different penalized estimators shrinking $\tilde{p}$ closer to $0$.}
	\label{fig:pen_sceme}
\end{figure}

All four of the penalized solutions above corresponding to some notion of success probabilities being ``close to $0$'' or ``not too large.'' Figure \ref{fig:pen_sceme} shows a schematic representation of the behavior of these estimators as a function of $\log(\lambda)$.

\subsection{Other Shrinkage Configurations}

The penalized estimators of Section 2.2 all revolve around the goal of ensuring that the estimates $\tilde{p}_i$ are close to $0$. If, on the other hand, it was desired to have estimates $\tilde{p}_i$ close to $1$, then by symmetry all of the examples considered could replace the $p_i$ in each of the penalty functions by $1-p_i$. Of course, many other types of penalties could also be of interest. For instance, consider the hypothetical example where a subject-matter expert expresses confidence that all of the $p_i$ should be close to some specified value $\kappa \in (0,1)$. For this specified $\kappa$, define
\[\mathrm{Pen}_{5}(\bm{p}|\kappa) = -\sum_{i=1}^{I}\left[\kappa \log(p_i) + (1-\kappa)\log(1-p_i)\right].\] This penalty function has a minimum when all the $p_i$ are equal to $\kappa$, and is unbounded above whenever one of the $p_i$ approach either $0$ or $1$. This penalty therefore shrinks the $p_i$ towards the specified $\kappa$ value. The penalized estimators are $$\tilde{p}_i = \frac{N_i}{N_i + \lambda}\ \hat{p}_i + \frac{\lambda}{N_i + \lambda}\ \kappa,\ i=1,\ldots,I.$$
For the $i^{th}$ variable, this estimator is a linear combination of the MLE and $\kappa$. The careful reader may also notice that this estimator has much in common with the Bayesian estimator of a binomial success probability with a beta prior. This example makes clear how the value of $\lambda$ controls whether the strength of evidence lies with the empirical estimator $\hat{p}_i$ or with the pre-specified reference $\kappa$. Similarly, say a subject-matter expert states that the success probabilities should all be ``close'' to one another, but without specifying a $\kappa$ value. For the WRI data, this is equivalent to requiring the $p_i$ to be near one another using some appropriate distance metric. For this, define the bounded penalty function
\[\mathrm{Pen}_{L_2}(\bm{p}) = \sum_{i=1}^{I}\sum_{j=1}^{I}\left(p_i-p_j\right)^2.\] Alternatively, if an unbounded penalty function is preferred, one could use
\[\mathrm{Pen}_{Q_2}(\bm{p}) = \sum_{i=1}^{I}\sum_{j=1}^{I}\left[\Phi^{-1}(p_i)-\Phi^{-1}(p_j)\right]^2\]
where $\Phi^{-1}$ is the standard normal quantile function. Neither of these penalties result in closed-form solutions for the shrinkage estimators $\tilde{p}_i$, $i=1,\ldots,I$.

\section{Data-Driven Shrinkage} \label{sec:CV}
In Section \ref{PenLik}, different penalty functions were considered for estimating $I$ independent binomial success probabilities assuming a known value of the shrinkage parameter $\lambda$. As $\lambda$ controls the relative importance of the penalty function, it is important to choose a value resulting in parameter estimates with small MSE. We present here how V-fold cross-validation (VFCV) can be used for selecting an optimal shrinkage parameter. While the VFCV approach is fully defined in this section, the interested reader can consult \cite{arlot2010survey} for a more in-depth discussion of this method as well as other cross-validation approaches.

Consider a dataset consisting of $I$ independently sampled variables, with the $i$th variable consisting of $n_i$ independent observations. Let $\bm{x}_i=(x_{i1},x_{i2},\ldots,x_{in_i})$ denote the observations corresponding to the $i$th variable. VFCV partitions the data into $V$ subsets of roughly equal size. For the $i$th variable, let $\mathcal{I}_{i,v}$, $v=1,\ldots,V$ denote a partition of the indices, such that $\bigcup_v\ \mathcal{I}_{i,v}=\{1,\ldots,n_i\}$ and $\mathcal{I}_{i,v_1}\bigcap\mathcal{I}_{i,v_2}=\varnothing$ for all $v_1\neq v_2$ with $v_1,v_2\in \{1,\ldots,V\}$. 

VFCV repeatedly creates subsets of the data for model training, in each instance leaving out one of the $V$ subsets per variable. The subsets left out in each iteration are then used for model validation. More specifically, the model building data subsets are used to estimate penalized parameter estimates for various degrees of penalty enforcement, say $M$ possible values of $\lambda$ satisfying $0=\lambda_1 < \lambda_2 < \cdots < \lambda_M$. The negative log-likelihood function for the validation data is then evaluated using penalized estimators corresponding to each possible value of $\lambda$. The optimal value $\lambda_{opt}$ is chosen to be the minimizer of the negative log-likelihood function averaged over the validation subsets.

Algorithmically, implementation of VFCV proceeds as follows:
\begin{itemize}
\item For the $i^{th}$ variable, form a training dataset by excluding the $v$th fold, $\bm{x}_{train,i}^{(v)}=\{x_{ij}:\ j\not\in \mathcal{I}_{i,v}\}$, and let the $v$th fold equal to the validation set, $\bm{x}_{valid,i}^{(v)}=\{x_{ij}:\ j \in \mathcal{I}_{i,v}\}$. Let $n_i^{(v)}$ denote the number of observations in $\bm{x}_{train,i}^{(v)}$. Also let $\bm{x}_{train}^{(v)}$ and $\bm{x}_{valid}^{(v)}$ denote the collection of the training and validation sets for all $I$ variables.
\item For each value $0=\lambda_0<\lambda_1<\ldots<\lambda_M$, find the estimators $\tilde{\bm{\theta}}_{train}^{(v)}(\lambda_m)$ that minimize the penalized negative log-likelihood function
\[D_k(\bm{\theta}) = - l\left(\bm{\theta}\left|\bm{x}_{train}^{(v)}\right.\right) + \lambda_m \bar{n}^{(v)} \mathrm{Pen}(\bm{\theta})\] where $\bar{n}^{(v)} = (1/I)\sum_i n_i^{(v)}$.
\item Calculate the validation function by evaluate the negative log-likelihood at this estimator,
\[\tilde{D}^{(v)}(\lambda_m) = -\ell\left(\left.\tilde{\bm{\theta}}_{train}^{(v)}(\lambda_m)\right|\bm{x}_{valid}^{(v)}\right).\]
\end{itemize}
The above bullets are repeated for $v=1,\ldots,V$ and the VFCV score is defined as
\begin{equation}
    \mathrm{CV}_m = \mathrm{CV}(\lambda_m) = \sum_{v=1}^{V} \tilde{D}^{(v)}(\lambda_m). \label{CVscore}
\end{equation}
The optimal shrinkage level is taken to be the minimizer of $\mathrm{CV}_m$, i.e. $\lambda_{opt} = \lambda_{m^\ast}$ with $m^\ast = \mathrm{argmin}_m \mathrm{CV}_m$.
Note that after the optimal penalty level has been chosen using VFCV, penalized estimators are calculated one more time using the full dataset. The penalized likelihood estimator with data-driven shrinkage, denoted $\bm{\tilde{\theta}}_{pen}$, is the minimizer of \[D_{opt}(\bm{\theta}) = -\ell(\bm{\theta}|\bm{x}) + \lambda_{opt}\, \bar{n}\, \mathrm{Pen}(\bm{\theta}) \]
where $\bar{n}=(1/I)\sum_i n_i$.
The literature on cross-validation recommends various choices for $V$, with common values ranging from $V=2$ to $V=10$. The choice $V=n$ is equivalent to leave-one-out cross-validation and can become computationally expensive. As discussed in \cite{arlot2010survey}, the size of the validation set has an effect on the bias of the penalized estimator, while the number of folds $V$ controls for the variance of the estimated penalization parameter. These authors also discuss some asymptotic considerations of cross-validation. If $n_{train}$ denotes the size of the training set, then for $n_{train}/n \rightarrow 1$, cross-validation is asymptotically equivalent to Mallows’ $C_p$ and therefore asymptotically optimal. Furthermore, if $n_{train}/n \rightarrow \gamma \in (0,1)$, then asymptotically the model is equivalent to Mallows' $C_p$ multiplied by (or over-penalized by) a factor $(1+\gamma)/(2\gamma)$. Asymptotics notwithstanding, throughout the remainder of this paper, an approach of $V=10$ is used. This strikes a balance between having larger training sets and reasonable computational costs.

\section{Simulation Studies}

In Section \ref{PenLik}, various shrinkage estimators for the binomial distribution were considered. Of course, the binomial model is not the only count model of interest. In this section, shrinkage estimation is considered for the binomial model as well as two related models, the zero-inflated binomial and the beta-binomial. In most scenarios investigated here, no closed-form solutions for the penalized estimators are available. Even so, these simulation studies are very useful for investigating the properties of different penalty functions and how they impact parameter estimation for the three models. Simulations are restricted to $I=10$ independent variables (passages), consisting of $N_i=N=40$ trials (passage length) and having $n_i=n=50$ independent observations (students) for $i=1,\ldots,I$. This choice was motivated in large part by the structure of the real data considered in this paper.

\subsection{The Binomial Model}
\begin{figure}[H]
	\begin{center}
		\includegraphics[trim={2cm 7cm 1.05cm 8cm},clip,width=9cm]{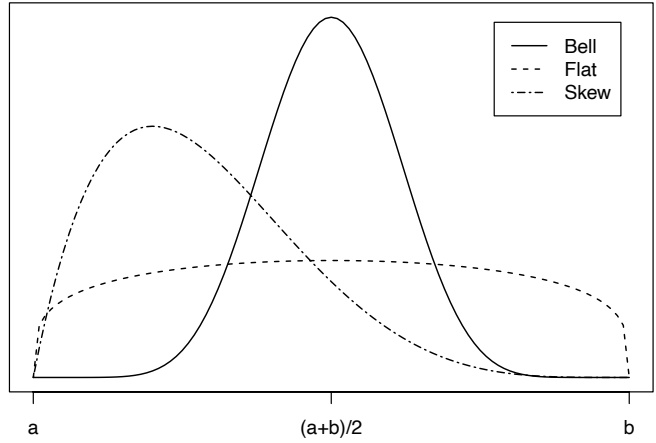}
		\caption{Success probability distributions considered in the simulation study.}
		\label{fig:SimParms}
	\end{center}
\end{figure}

In the simulation, samples $\mathcal{X} = \{X_{ij}, i=1,\ldots,I,\ j = 1,\ldots,n\}$ were generated with independent observations $X_{ij} \sim \mathrm{Bin}(N,p_i)$ and $(I,N,n)=(10,40,50)$. The binomial success probabilities $p_i$ were sampled from a scaled beta distribution. Three shapes of the success probability distribution were considered, namely a skewed distribution $(p_i-a)/(b-a) \sim \mathrm{Beta}(2, 5)$, a very flat distribution $(p_i-a)/(b-a) \sim \mathrm{Beta}(5/4, 5/4)$, and a bell-shaped distribution $(p_i-a)/(b-a) \sim \mathrm{Beta}(10, 10)$. The three success probability distributions are illustrated in Figure \ref{fig:SimParms}. When considering shrinkage to $0$, we chose scaling parameters $(a,b)\in \{(0.01,0.05),(0.01,0.10),(0.30,0.50)\}$ and when considering shrinkage closer to one another, we chose $(a,b)\in \{(0.01,0.05),(0.08,0.20),(0.31,0.35)\}$. In total, this makes for $18$ simulation configurations: $3$ distributions for the $p_i$ $\times$ $2$ types of shrinkage $\times$ $3$ choices of $(a,b)$ for each shrinkage type. The $\lambda$ term controlling how aggressively the penalty gets enforced was chosen using cross-validation using $63$ possible values ranging from $0$ to $10,000$ spaced approximately equidistant on a logarithmic scale. These $\lambda$ values were selected (after some trial-and-error) to ensure they cover the spectrum of negligible penalization ($\lambda=0$) through the penalty dominating ($\lambda=10,000$). VFCV was used to choose the optimal $\lambda$ for each simulated dataset. In addition to the penalized estimators, maximum likelihood estimators were also calculated. In total, $K=500$ samples were generated for each of the $18$ simulation configurations. 

Summarized in the tables below are the Monte Carlo estimates of the MSE ratios. For the $k$th sample $\mathcal{X}_k$, let $\bm{p}_k=\left(p_{k,1},\ldots,p_{k,10}\right)$ denote the true success probabilities simulated from a specified scaled Beta distribution. Let $\hat{\bm{p}}_k$ denote the MLE and let $\tilde{\bm{p}}_k$ denote a penalized estimator found using VFCV. Define Sum of Squared Deviations $\mathrm{SSD}(\bm{p}_1,\bm{p}_2) = \sum_{i=1}^{I} (p_{1i} - p_{2i})^2$. 
The Monte Carlo MSE ratios are subsequently defined as 
\[\mathrm{MSE}_{\mathrm{Pen}} =  \frac{(1/K)\sum_{k=1}^{K}\mathrm{MSD}(\tilde{\bm{p}}_k,\bm{p}_k)}{(1/K)\sum_{k=1}^{K}\mathrm{MSD}(\hat{\bm{p}}_k,\bm{p}_k)} \]
where the subscript ``$\mathrm{Pen}$'' emphasizes the specific penalty function used to obtain the estimators. Maximum likelihood is often considered a ``gold standard'' estimation method. Therefore, we do not report the estimated MSE values themselves, but rather emphasize the MSE ratios comparing the penalized estimators to maximum likelihood. An MSE ratio less than $1$ indicates superior performance of the penalized estimator, while an MSE ratio exceeding $1$ indicates that the unpenalized estimator is preferred.

In Table \ref{tab:Bin_to_zero}, the results of shrinking success probabilities to zero are presented using the penalties $\mathrm{Pen}_j(\bm{p})$, $j=1,\ldots,4$. In Table \ref{tab:Bin_closer}, the results of shrinking success probabilities closer to one another using penalties $\mathrm{Pen}_{L_2}$ and $\mathrm{Pen}_{Q_2}$ are presented. To recall these penalties, consult Section 2.2 of this paper. The tables also report summary measures for the count variables simulated under the different configurations, taking $\bar{\mathrm{E}}(X) = (1/I)\sum_{i=1}^{I}\mathrm{E}(X_i)$ and $\bar{\mathrm{S}}(X) = \left[(1/I)\sum_{i=1}^{I}\mathrm{Var}(X_i)\right]^{1/2}$ as summary measures of location and spread.

\begin{table}[H]
\tbl{{MSE ratios comparing penalized parameter estimates to maximum likelihood when shrinking estimators to 0.}}
   { \begin{tabular}{|c|c|c|c|c|c|c|c|}
    \cline{5-8}
        \multicolumn{4}{c|}{} & \multicolumn{4}{c|}{Penalty} \\ \hline
        $p_i \in(a,b)$ & Shape & $\bar{\mathrm{E}}(X)$ & $\bar{\mathrm{S}}(X)$ & Pen$_1$ & Pen$_2$ & Pen$_3$ & Pen$_4$  \\ \hline
          $(0.01,0.05)$ & Skew & 0.857 & 0.950 & 0.999 & 0.956 & 0.988 & 1.382 \\ 
         ~ & Flat & 1.200 & 1.158 & 0.999 & 0.968 & 0.995 & 1.012  \\ 
         ~ & Bell & 1.200 & 1.093 & 0.999 & 0.961 & 0.997 & 1.011  \\ \hline
         $(0.01,0.10)$ & Skew & 1.429 & 1.303 & 0.999 & 0.977 & 0.995 & 1.013 \\ 
         ~ & Flat & 2.200 & 1.726 & 1.000 & 0.982 & 0.994 & 1.004  \\ 
         ~ & Bell & 2.200 & 1.493 & 1.000 & 0.978 & 0.996 & 1.002  \\ \hline
         $(0.30,0.50)$ & Skew & 14.286 & 3.283 & 0.998 & 0.998 & 1.015 & 0.999 \\ 
         ~ & Flat & 16.000 & 3.749 & 0.999 & 0.998 & 1.037 & 1.000    \\ 
         ~ & Bell & 16.000 & 3.216 & 1.000 & 0.999 & 1.027 & 1.003  \\ \hline
    \end{tabular} }
    \label{tab:Bin_to_zero}
\end{table}

\begin{table}[H]
\tbl{MSE ratios comparing penalized parameter estimates to maximum likelihood when shrinking estimators closer to one another.}
    {\begin{tabular}{|c|c|c|c|c|c|c|}
    \cline{5-6}
        \multicolumn{4}{c|}{} & \multicolumn{2}{c|}{Penalty} 
        \\  \hline
         $p_i \in(a,b)$ & Shape & $\bar{\mathrm{E}}(X)$ & $\bar{\mathrm{S}}(X)$ & L$_2$ & Q$_2$ \\ \hline
         $(0.01,0.05)$ & Skew & 0.857 & 0.950 & 0.928 & 0.906  \\ 
         ~ & Flat & 1.200 & 1.159 & 0.935 & 0.942  \\ 
         ~ & Bell & 1.200 & 1.093 & 0.705 & 0.704 \\ \hline
         $(0.08,0.20)$ & Skew & 4.571 & 2.149 & 0.960 & 0.952 \\ 
         ~ & Flat & 5.600 & 2.533 & 0.969 & 0.973  \\ 
         ~ & Bell  & 5.600 & 2.255 & 0.854 & 0.856\\ \hline
         $(0.31,0.35)$ & Skew & 12.857 & 2.965 & 0.411 & 0.411  \\ 
         ~ & Flat & 13.200 & 3.003 & 0.652 & 0.652  \\ 
         ~ & Bell & 13.200 & 2.979 & 0.292 & 0.293  \\ \hline
    \end{tabular}}
    \label{tab:Bin_closer}
\end{table}

In Table \ref{tab:Bin_to_zero}, the best-performing penalty function when shrinking to $0$ is $\mathrm{Pen}_2(\bm{p}) = \sum_i p_i^2$. Even so, the relative improvement in efficiency is small throughout. The only penalty that consistently leads to worse performance than maximum likelihood is $\mathrm{Pen}_4(\bm{p})$. Recall that this penalty function is not associated with a norm and is able to very aggressively shrink success probabilities to $0$. This simulation suggests that, at least in the scenarios considered, this penalty shrinks too aggressively. For the other three estimators, VFCV results in penalized estimators with slightly better performance than MLE. 

In Table \ref{tab:Bin_closer}, the performance of the $L_2$ and $Q_2$ penalties is nearly indistinguishable. When shrinking parameters closer to one another, large gains in efficiency are sometimes realized. This is especially notable when the Beta shape from which the success probabilities are generated is bell-shaped, i.e. the $p_i$ are close to one another. In all instances, VFCV results in penalized estimators with performance superior to maximum likelihood. Altogether, these simulations illustrate that both the average success probability and the spacing of the $p_i$  relative to that average are important in determining the reduction in MSE. In Table 2, we also note that the MSE ratio tends to decrease, indicating better efficiency, when $\bar{\mathrm{E}}(X)$ is further from $0$. For penalties shrinking the $p_i$ closer to one another,  an MSE ratio below $0.3$ was realized, showing dramatic improvement due to shrinkage. 

\subsection{The Zero-inflated Binomial Distribution}
The probability mass function of the zero-inflated binomial (ZIB) distribution is
\[f(x|N,\pi,\gamma) = \left\{ 
\begin{array}{ll}
	\gamma + (1 - \gamma)(1 - \pi)^{N} & \text{for $x = 0$} \\
  (1 - \gamma)\dbinom{N}{x}\pi^{x}(1-\pi)^{N-x} & \text{for $x = 1,\ldots, N$}
\end{array}%
\right. \]
where $\gamma$ represents the excess zero probability, and $\pi$ and $N$ are the binomial success probability and number of trials. For $X\sim \mathrm{ZIB}(N,\pi,\gamma)$, it follows that $E[X] = N\pi(1-\gamma)$. Consequently, we note the overall expected success proportion in a ZIB is $p = E[X]/N = \pi(1 - \gamma)$. The parameter $p$ is of primary interest when considering possible penalty functions, especially under the assumption that the different ZIB distributions are ``similar'' to one another.

In the simulation study, samples $\mathcal{X} = \{X_{ij}, i=1,\ldots,I,\ j = 1,\ldots,n\}$ were generated with independent ZIB variables, $X_{ij} \sim \mathrm{ZIB}(N, \pi_i, \gamma_i)$ and $(I,N,n) = (10,40,50)$. The overall success proportions $p_i$ and the excess zero probabilities $\gamma_i$ were sampled from the scaled beta distributions as per Figure \ref{fig:SimParms} with the specific bounds $(a_1,b_1)$ for the $p_i$ and $(a_2,b_2)$ for the $\gamma_i$ listed in the table below. In total, $12$ simulation configurations were considered: $3$ distributional shapes $\times$ $4$ choices for $(a_1,b_1,a_2,b_2)$. In the simulations, the binomial success probabilities $\pi_i$ were recovered from the $p_i$ and $\gamma_i$ through $\pi_i = p_i/(1-\gamma_i)$, $i=1,\ldots,I$. A total of $k=500$ samples were simulated under each configuration. 

The ZIB simulation considered three penalty functions, $\mathrm{Pen}_2(\bm{p})=\sum_i p_i^2$, $\mathrm{Pen}_{L_2}(\bm{p})=\sum_i\sum_j (p_i-p_j)^2$, and $\mathrm{Pen}_{full}(\bm{\gamma},\bm{\pi}) = \sum_{i} \sum_{j} (\gamma_i - \gamma_j)^2 + \sum_{i} \sum_{j} (\pi_i - \pi_j)^2$. The first of these, termed \textit{zero shrinkage}, results in estimated $p_i$ closer to $0$. The second, termed \textit{mean shrinkage}, results in $p_i$ closer to one another. The third, termed \textit{full shrinkage}, shrinks all $\gamma_i$ closer to one another and all $\pi_i$ closer to one another. While both the penalties $\mathrm{Pen}_{L_2}$ and $\mathrm{Pen}_{full}$ have the goal of estimating models that are ``similar'' to one another, the second penalty is much more strict. To see this, consider two passages with equal average difficulty $p_i=p_j$. Under the first penalty, the contribution of their squared difference is $0$. However, it is possible to have $(\gamma_i,\pi_i)\neq(\gamma_j,\pi_j)$ even when $p_i=p_j$, meaning there could conceivably be a non-zero contribution to the full shrinkage penalty function.

In addition to using VFCV to select the level of shrinkage for the above three penalities, a combined estimator, termed \textit{minCV}, was calculated by selecting among the three penalized estimators the one with the smallest VFCV score. The same set of $63$ $\lambda$ values ranging from $0$ to $10,000$ were used. The Monte Carlo MSE ratios for the success proportions $\bm{p}$ are in Table \ref{tab:ZIB_MSE}. The MSE ratios for $\bm{\gamma}$ and $\bm{\pi}$ were also calculated, and these can be found in Table 8 of the Supplemental Material.

\begin{table}[H]
\tbl{MSE ratios for ZIB success proportions $\bm{p}=(p_1,\ldots,p_{10})$ comparing penalized parameter estimates to maximum likelihood for different penalization approaches.}
    {\begin{tabular}{|c|c|c|c|c|c|c|c|c|}
    \cline{6-9}
        \multicolumn{5}{c|}{} & \multicolumn{4}{c|}{Penalty} \\ \hline
        $\pi_i \in (a_1,b_1)$ & $\gamma_i \in (a_2,b_2)$ & Shape & $\bar{\mathrm{E}}(X)$ & $\bar{\mathrm{S}}(X)$ & Zero & Mean & Full & minCV \\ \hline
        $(0.01,0.05)$ & $(0.10,0.14)$ & Skew & 0.761 & 0.935 & 0.957 & 0.888 & 0.981 & 0.958   \\ 
        ~ & ~ & Flat & 1.055 & 1.153 & 0.977 & 0.942 & 0.979 & 0.983  \\ 
        ~ & ~ & Bell & 1.056 & 1.097 & 0.964 & 0.668 & 0.836 & 0.755  \\ \hline
        $(0.04,0.06)$ & $(0.20,0.30)$ & Skew & 1.410  & 1.395 & 0.968 & 0.364 & 0.368 & 0.356 \\ 
        ~ & ~ & Flat & 1.502 & 1.485 & 0.971 & 0.562 & 0.526 & 0.523  \\ 
        ~ & ~ & Bell & 1.496 & 1.477 & 0.968 & 0.258 & 0.246 & 0.239  \\ \hline
        $(0.15,0.30)$ & $(0.04,0.06)$ & Skew & 7.364 & 3.064 & 1.006 & 0.969 & 0.860 & 0.885  \\ 
        ~ & ~ & Flat  & 8.551 & 3.586 & 1.010 & 1.005 & 0.808 & 0.819 \\ 
        ~ & ~ & Bell & 8.552 & 3.296 & 1.009 & 0.821 & 0.873 & 0.899  \\ \hline
        $(0.05,0.06)$ & $(0.20,0.70)$ & Skew & 1.389 & 1.526 & 0.963 & 0.203 & 0.635 & 0.273  \\ 
        ~ & ~ & Flat & 1.209 & 1.531 & 0.955 & 0.223 & 0.934 & 0.259  \\ 
        ~ & ~ & Bell & 1.210 & 1.529 & 0.951 & 0.183 & 0.372 & 0.245 \\ \hline
    \end{tabular}}
    \label{tab:ZIB_MSE}
\end{table}

Consider now Table \ref{tab:ZIB_MSE}. While \textit{zero shrinkage} does result in some efficiency gains in most scenarios, overall MSE ratios close to $1$ suggest little improvement from using this penalty. On the other hand, both \textit{mean} and \textit{full shrinkage} result in large decreases in the MSE ratios. Overall, it cannot be said that either \textit{mean} and \textit{full} shrinkage performs best. This makes sense, as it depends on the configuration of all parameters and not just the mean parameters. Finally, while \textit{minCV} does not always have the smallest MSE ratio, it is generally close to the minimum. This suggests that data-driven selection of the level of shrinkage \textit{as well as} the penalty function leads to good performance for the model.

\subsection{The Beta-Binomial Model}
The probability mass function of the beta-binomial distribution is given by
\[ f(x|N,\alpha,\beta) = \binom {N}{x}\frac{B(x+\alpha,N - x + \beta)}{B(\alpha,\beta)} ,\ x = 0,1,...,N\]
where $B(x,y) = \int_0^1 t^{x-1}(1-t)^{y-1}dt$ is the so-called Beta function, $N$ is the number of trials, and $\alpha>0$ and $\beta>0$ control the mean and variance of the distribution. Defining $p = \alpha/(\alpha + \beta)\in (0,1)$ and $\nu = (\alpha+\beta+N)/(\alpha+\beta+1)\in (1,N)$, the mean and variance of the distribution can be written as $E[X]=Np$ and $Var[X] = Np(1-p)\nu$. In this parameterization, $p$ and $\nu$ denote, respectively, the expected success proportion successes and the the over-dispersion relative to a binomial distribution with the same mean value. 

Samples $\mathcal{X} = \{X_{ij}, i=1,\ldots,I,\ j = 1,\ldots,n\}$ were generated with independent Beta-Binomial variables, $X_{ij} \sim \mathrm{BetaBin}(N, \alpha_i, \beta_i)$, with $(I,N,n)=(10,40,50)$. The overall success proportions $p_i$ and the overdispersion measures $\nu_i$ were sampled from the scaled beta distributions as per Figure \ref{fig:SimParms} with the specific bounds $(a_1,b_1)$ for the $p_i$ and $(a_2,b_2)$ for the $\nu_i$ listed in Table \ref{tab:BBin_MSE}. Again, $12$ simulation configurations were considered. In the simulation, parameters $\alpha_i$ and $\beta_i$ for the beta-binomial distribution were recovered from the simulated $p_i$ and $\nu_i$ through the relationships in the preceding paragraph. A total of $K=500$ samples were simulated under each configuration.

As in Section 4.2, three penalty functions. Letting $p_i = \alpha_i/(\alpha_i+\beta_i)$, $i=1,\ldots,I$, these were $\mathrm{Pen}_2(\bm{p})=\sum_i p_i^2$, $\mathrm{Pen}_{L_2}(\bm{p})=\sum_i\sum_j (p_i-p_j)^2$, and $\mathrm{Pen}_{full}(\bm{\alpha},\bm{\beta}) = \sum_{i} \sum_{j} (\alpha_i - \alpha_j)^2 + \sum_{i} \sum_{j} (\beta_i - \beta_j)^2$. These are again termed, respectively, \textit{zero shrinkage}, \textit{mean shrinkage}, and \textit{full shrinkage}. In addition to the three penalized estimators, an estimator termed \textit{minCV} was calculated by selecting among the three penalized estimators the one with the smallest CV score. The MSE ratios for all estimators are reported in Table \ref{tab:BBin_MSE}. The table shows the results for the success proportions $\bm{p}$, and the equivalent results for $\bm{\alpha}$ and $\bm{\beta}$ can be found in Table 9 of the Supplemental Material.

\begin{table}[H]
\tbl{MSE ratios for Beta-Binomial success proportions $\bm{p}=(p_1,\ldots,p_{10})$ comparing penalized parameter estimates to maximum likelihood for different penalization approaches.}
{\begin{tabular}{|c|c|c|c|c|c|c|c|c|}
    \cline{6-9}
        \multicolumn{5}{c|}{} & \multicolumn{4}{c|}{Penalty} \\ \hline
        $p_i \in (a_1,b_1)$ & $\nu_i \in (a_2,b_2)$ & Shape & $\bar{\mathrm{E}}(X)$ & $\bar{\mathrm{S}}(X)$ & Zero & Mean & Full & minCV \\ \hline
        $(0.05, 0.10)$ & $(4, 6)$ & Skew & 2.361 & 3.102 & 0.917 & 0.474 & 0.428 & 0.429  \\ 
        & ~ & Flat & 2.733 & 3.480 & 0.928 & 0.702 & 0.591 & 0.604 \\ 
        & ~ & Bell & 2.730 & 3.455 & 0.921 & 0.290 & 0.270 & 0.271  \\ \hline
        $(0.12, 0.22)$ & $(2, 5)$ & Skew & 5.742 & 3.755 & 0.974 & 0.722 & 0.726 & 0.708  \\
        & ~ & Flat & 6.513 & 4.419 & 0.977 & 0.903 & 0.948 & 0.889  \\
        & ~ & Bell & 6.515 & 4.327 & 0.973 & 0.476 & 0.466 & 0.463  \\ \hline
        $(0.17, 0.22)$ & $(3, 8)$ & Skew & 6.947 & 4.929 & 0.971 & 0.301 & 0.400 & 0.331  \\ 
        & ~ & Flat & 7.230 & 5.557 & 0.971 & 0.445 & 0.762 & 0.481  \\
        & ~ & Bell & 7.221  & 5.555 & 0.968 & 0.217 & 0.242 & 0.227  \\ \hline
        $(0.05, 0.06)$ & $(2, 10)$ & Skew & 1.952 & 2.723 & 0.905 & 0.170 & 0.469 & 0.211  \\ 
        & ~ & Flat & 1.943 & 3.139 & 0.891 & 0.188 & 0.733 & 0.213  \\
        & ~ & Bell & 1.949 & 3.183 & 0.893 & 0.155 & 0.187 & 0.175  \\ \hline
    \end{tabular}}
    \label{tab:BBin_MSE}
\end{table}

Inspecting Table \ref{tab:BBin_MSE}, \textit{zero shrinkage} is noted to be the least effective approach here, even while still being more effective than maximum likelihood. For most of the simulation configurations, MSE ratios under \textit{mean} and \textit{full shrinkage} are comparable. Here, the \textit{minCV} approach is also very impressive, in most instances nearly matching the best-performing method. This reaffirms that VFCV can be effectively used to choose both the level of shrinkage for a specific penalty function, but then also choose from among competing penalty functions.

\section{Data Analysis}
\begin{figure}[H]
	\begin{center}
		\includegraphics[width=10cm]{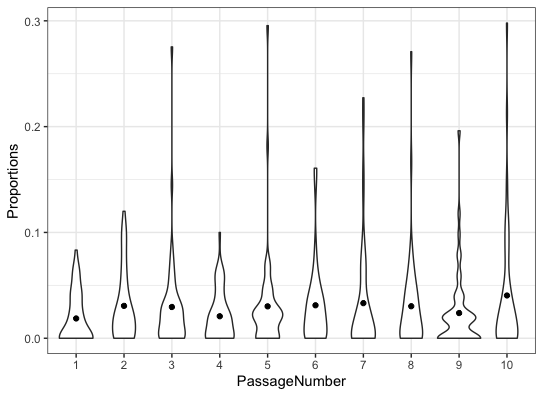}
		\caption{WRI proportions for the ten passages}
		\label{fig:WRIProportions}
	\end{center}
\end{figure}

The methodology developed in this paper was motivated by the oral reading fluency data collected from a sample of $508$ elementary-school aged children. Each child was randomly assigned one of ten available passages to read. This resulted in around 50 Words Read Incorrectly (WRI) scores per passage. Table \ref{SumStats} reports specific details for passage length, sample size per passage, as well as the minimum, median, and maximum WRI scores. Of interest is to accurately and efficiently estimate passage difficulty as measured by the average proportion of words read incorrectly. Note that higher WRI proportions (i.e. WRI counts divided by passage length) indicate that a passage is more difficult. Figure \ref{fig:WRIProportions} provides information about the passage-specific WRI proportions. The solid dot in each violin plot represents the mean WRI proportion. The means correspond to the unpenalized maximum likelihood estimates of passage difficulty.  

\begin{table}[H]
\tbl{Passage-level summary statistics}
    {\begin{tabular}{|c|c|c|c|c|c|c|c|c|c|c|}
    \hline
        Passage Number & 1 & 2 & 3 & 4 & 5 & 6 & 7 & 8 & 9 & 10 \\ \hline
        Sample Size & 49 & 51 & 51 & 50 & 52 & 51 & 50 & 53 & 51 & 50\\ \hline
        Passage Length & 48 & 50 & 69 & 50 & 44 & 56 & 44 & 48 & 51 & 47 \\ \hline
        Minimum WRI & 0 & 0 & 0 & 0 & 0 & 0 & 0 & 0 & 0 & 0 \\ \hline
        Median WRI & 0 & 1 & 1 & 1 & 1 & 1 & 1 & 1 & 1 & 1 \\ \hline
        Maximum WRI & 4 & 6 & 19 & 5 & 13 & 9 & 10 & 13 & 10 & 14 \\ \hline
    \end{tabular}}
    \label{SumStats}
\end{table}

The mean WRI proportions in Figure \ref{fig:WRIProportions} appear fairly close to one another, supporting the assumption that the passages fall within a narrow range of difficulty. Thus, it is plausible that appropriate shrinkage will result in improved estimates of difficulty.

Three models and three types of shrinkage were considered for the data at hand. We remind the reader that classic selection criteria such as AIC and BIC cannot easily be applied in parameter shrinkage settings unless one is able to calculate the effective number of parameters. In a penalized model with $K$ specified parameters, the \textit{effective number of parameters} $\tilde{K}$ can be dramatically smaller than $K$. Generally, there is no easy way to calculate $\tilde{K}$ in penalized models. We therefore used cross-validation (CV) to select the best model, noting that such CV scores as per \citet{geisser1975predictive} represent a \textit{discrepancy measure} for each model. The lowest CV score corresponds to the smallest empirical discrepancy between observed data and estimated model. Therefore, the smallest CV score corresponds to the optimal model choice. In each model under consideration, the same set of data partitions was used to select a smoothing parameter with VFCV with $V=10$ fold. Table \ref{tab:data_analysis_CV} reports the VFCV scores as defined in \eqref{CVscore}. When the penalty in the table is specified as ``None'', the VFCV score corresponds to the unpenalized maximum likelihood estimators.

\begin{table}[H]
\tbl{10-fold CV scores and optimal $\lambda$ values for the three distributions considered.}
    {\begin{tabular}{|c|c|c|c|}
    \hline
        Distribution & Penalty & VFCV & $\log(\lambda_{opt}+1)$ \\ \hline
        Binomial & None & $1025.5$ & -- \\
        & Zero & $1024.9$ & $3.56$ \\
         & Mean & $1017.1$ & $4.36$ \\ \hline
        ZIB & None & $964.7$ & -- \\
        & Zero & $964.3$ & $2.78$ \\
        & Mean & $959.6$ & $3.96$ \\ 
         & Full & $950.4$ & $3.56$ \\ \hline
        BetaBin & None & $869.7$ & -- \\
        & Zero &  $869.5$ & $2.41$ \\
        & Mean &  $866.3$ & $3.56$ \\
         & Full & $851.9$ & $0.04$ \\ \hline
    \end{tabular}}
    \label{tab:data_analysis_CV}
\end{table}

\begin{table}[H]
\tbl{Beta-binomial parameter estimates for the WRI data.}
   {\begin{tabular}{|c|c|c|c|c|c|c|c|c|c|}
    \cline{2-10}
        \multicolumn{1}{c|}{} & \multicolumn{3}{c|}{Maximum Likelihood} & \multicolumn{3}{c|}{Mean Shrinkage} & \multicolumn{3}{c|}{Full Shrinkage} \\ \hline
         Passage & $\hat\alpha$ & $\hat\beta$ & $\hat{p}$ & $\tilde\alpha$ & $\tilde\beta$ & $\tilde{p}$ & $\tilde\alpha$ & $\tilde\beta$ & $\tilde{p}$ \\ \hline
        P1 & 1.28 & 66.34 &  0.019 & 1.20 & 52.67 & 0.022 & 0.70 & 27.45 & 0.025 \\ \hline
        P2 & 1.51 & 45.15 & 0.032 & 1.50 & 47.47 & 0.031 & 0.91 & 27.44 & 0.032 \\ \hline
        P3 & 0.84 & 19.85 & 0.040 & 0.80 & 22.81 & 0.034 & 0.96 & 27.44 & 0.034 \\ \hline
        P4 & 2.47 & 160.0 & 0.015 & 2.25 & 123.5 & 0.018 & 0.67 & 27.45 & 0.024 \\ \hline
        P5 & 1.17 & 42.54 & 0.027 & 1.17 & 41.65 & 0.027 & 0.83 & 27.45 & 0.030 \\ \hline
        P6 & 1.18 & 29.10 & 0.039 & 1.13 & 32.52 & 0.034 & 0.97 & 27.44 & 0.034 \\ \hline
        P7 & 0.53 & 19.48 & 0.026 & 0.53 & 18.75 & 0.027 & 0.74 & 27.44 & 0.026 \\ \hline
        P8 & 0.87 & 25.37 & 0.033 & 0.86 & 27.20 & 0.031 & 0.88 & 27.44 & 0.031 \\ \hline
        P9 & 0.85 & 32.25 & 0.026 & 0.84 & 30.74 & 0.027 & 0.79 & 27.45 & 0.028 \\ \hline
        P10 & 0.65 & 17.03 & 0.037 & 0.63 & 19.14 & 0.032 & 0.89 & 27.44 & 0.031 \\ \hline
    \end{tabular} }
    \label{tab:BetaBinParms}
\end{table}

It is evident from Table \ref{tab:data_analysis_CV} that all variations of the beta-binomial model have much lower cross-validation scores than either the binomial or zero-inflated binomial models. Furthermore, VFCV never selects unpenalized maximum likelihood model for any of the distributions considered. With regards to penalty type, full shrinkage works best for this model with mean shrinkage being a distant second choice. Table \ref{tab:BetaBinParms} shows the beta-binomial parameter estimates obtained using maximum likelihood as well as penalized likelihood with mean shrinkage and full shrinkage. It is interesting to note that in the full shrinkage solution, the $\tilde{\beta}_i$ values have all been shrunk to within $0.01$ of a common value, but the $\tilde{\alpha}_i$ still exhibit a fair spread of values. For unpenalized maximum likelihood, the estimated success proportions range from $0.019$ to $0.04$, while the full shrinkage values range from $0.024$ to $0.034$. The latter shows much more adherence to the idea that the passages are similar in terms of difficulty.

\begin{figure}[H] 
    \centering{
\includegraphics[scale=0.9]{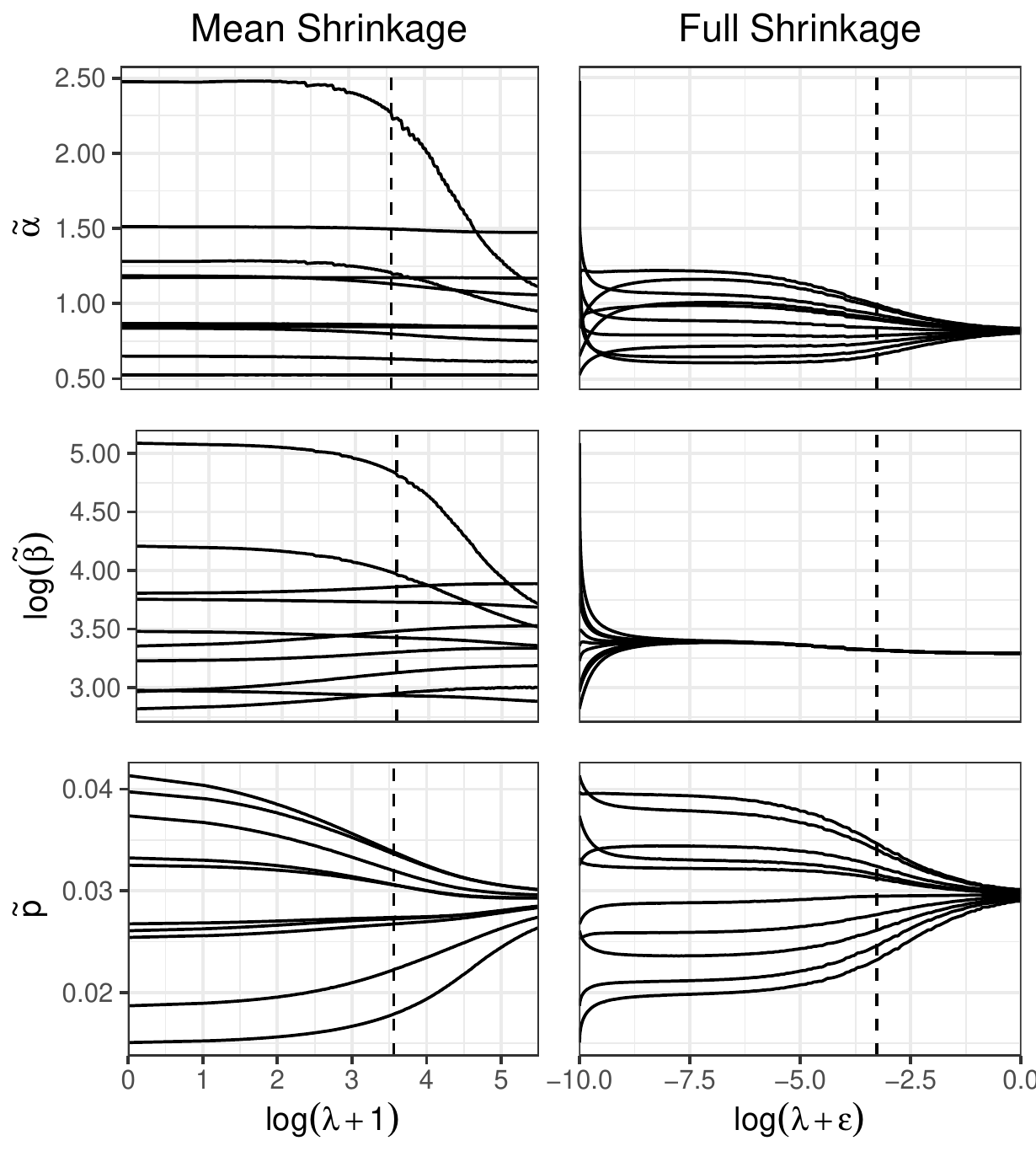}}
\caption{Beta-binomial parameter estimates under mean shrinkage and full shrinkage. Dashed line indicates optimal shrinkage. Scale value to improve full shrinkage plot readability is $\varepsilon=e^{-10}$.}
\label{fig:BetaBinParms}
\end{figure} 

For the interested reader, Figure \ref{fig:BetaBinParms} shows the penalized likelihood estimate trajectories for mean shrinkage and full shrinkage as a function of $\lambda$. The estimates of $\tilde{\beta}$ are presented on a logarithmic scale. For mean shrinkage, the horizontal scale is $\log(\lambda+1)$ and for full shrinkage it is $\log(\lambda+\varepsilon)$ with $\varepsilon=10^{-10}$. These adjustments were all made to improve readability of the plots. Dashed vertical lines indicate the optimal shrinkage solutions as determined by VFCV. 

Under mean shrinkage, the passage-specific $\tilde{\alpha}_i$ and $\tilde{\beta}_i$ still exhibit a large spread even when the success proportions $\tilde{p}_i= \tilde{\alpha}_i/(\tilde{\alpha}_i+\tilde{\beta}_i)$ are close to one another. Under full shrinkage, the $\tilde{\beta}_i$ values are very quickly shrunk to a nearly common value while the $\tilde{\alpha}_i$ still exhibit some spread. 

\begin{figure}[H] 
    \centering{
\includegraphics[trim={0.1cm 0.5cm 1.05cm 0.7cm},clip,scale=0.63]{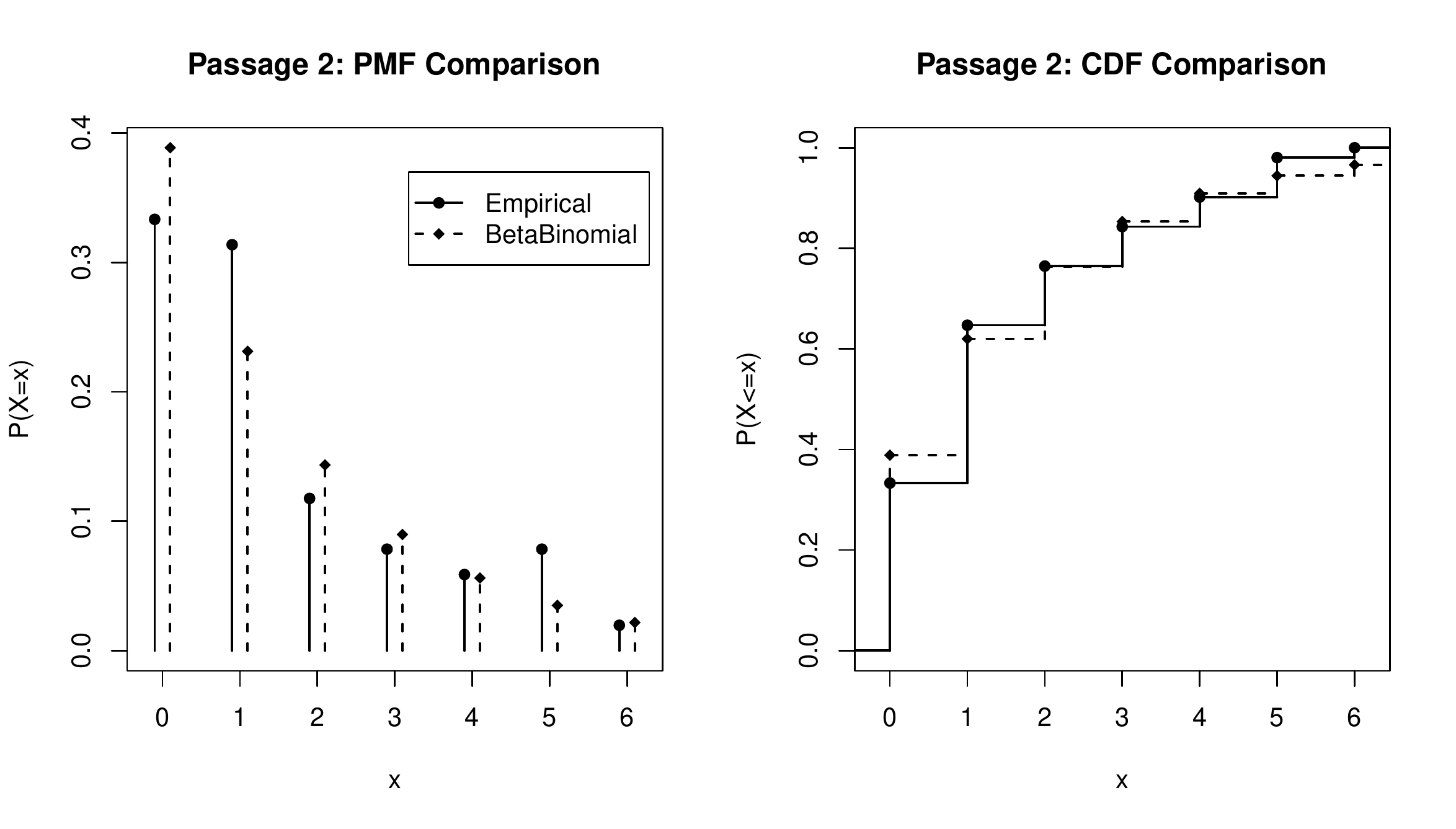}}
\caption{Empirical and penalized model-based pmf and cdf comparisons for the Passage 2 data.}
\label{fig:Pass2}
\end{figure}

One last matter that we will briefly address is that of post-selection model checking. Using VFCV above, the penalized beta-binomial model with full parameter shrinkage has been selected as the best model in a \textit{relative} sense. If one wishes to evaluate how well the model fits in an \textit{absolute} sense, one might compare the empirical and penalized model-based pmfs or cdfs. Figure \ref{fig:Pass2} shows both of these comparisons using the Passage 2 data as an example. These figures are presented with a note of caution -- the penalized model-based probabilities will almost never be as close to the empirical probabilities as the unpenalized probabilities based on the same parametric model and estimated for that specific passage only i.e. ignoring the data from other passages. As such, rather than a visual inspection, one may wish to use a more formal diagnostic tool. Pearson's chi-square goodness-of-fit statistic is one possibility worth considering. The use of this statistic is complicated by two matters. Firstly, as per \cite{chernoff1954use}, the Pearson statistic no longer has a limiting $\chi^2$ distribution when evaluated using estimated model parameters. Secondly, the effect of parameter penalization and model selection will further impact the distribution of the statistic. Therefore, to find sensible critical values, one would have to rely on a Monte Carlo procedure that incorporates both penalization and selection. This is a computationally burdensome procedure that we do not further consider in the present paper.

\section{Conclusions}

The goal of this project was defining and exploring penalized parameter estimators of passage difficulty from independent multivariate count data. WRI scores realized by $508$ students during an ORF assessment motivated the work and these data were analyzed in Section 5. The simulation results presented show that across the different count distributions and simulation configurations considered, large decreases in MSE relative to unpenalized maximum likelihood were often achieved. There is also very little risk in using penalized likelihood, as V-fold cross validation never resulted in a large increase in MSE. In fact, the \textit{minCV} approach explored in the simulations point the cross-validation being able to choose not just the appropriate level of shrinkage, but also the most appropriate penalty function under consideration. When applying the methodology to the observed WRI data, a penalized beta-binomial model is selected. This choice results in penalized estimators of the passage difficulty with a much tighter spread. This affirms the expectation that the passages are similar in difficulty, with estimated difficulty scores ranging from $2.4\%$ to $3.4\%$. Even so, this does highlight one important avenue for future research. If students are reading different passages to assess ORF, it is desirable to have a method that standardizes WRI scores to be independent of passage difficulty. In practice, students also typically read multiple passages, so exploring methods accounting for correlated WRI scores need to be considered in future.

\section*{Acknowledgements}

The research reported here was partially supported by the Institute of Education Sciences, U.S. Department of Education, through Grant R305D200038
to Southern Methodist University. The opinions expressed are those of the authors and do not represent
views of the Institute or the U.S. Department of Education.

\bibliography{references.bib}
\bibliographystyle{apalike}
\newpage
\section*{Supplemental Material}

In the main paper, simulation results are only presented for the success proportion estimates $\tilde{p}_i$. Of course, it is reasonable to assess the effect of shrinkage on other model parameters as well. The MSE ratios as defined in Section 3 are presented here when estimating the parameters $(\gamma_i,\pi_i)$ in the zero-inflated binomial model, see Table 7, and when estimating the parameters $(\alpha_i,\beta_i)$ in the beta-binomial model, see Table 8. These results confirm that shrinkage can be very beneficial to the estimation of incidental model parameters as well.

\begin{table}[H]
\tbl{MSE ratios for Zero-Inflated Binomial success proportions $\bm{\gamma}$ and $\bm{\pi}$ comparing penalized parameter estimates to maximum likelihood for different penalization approaches.}
   {\begin{tabular}{|c|c|c|c|c|c|c|c|c|c|c|}
    \cline{4-11}
        \multicolumn{3}{c|}{} & \multicolumn{4}{c|}{$\gamma_i$} & \multicolumn{4}{c|}{$\pi_i$}  \\ \hline
        $p_i \in (a_1,b_1)$  & $\gamma_i \in (a_2,b_2)$  & Shape & Zero & Mean & Full & Min CV & Zero & Mean & Full & Min CV \\ \hline
        (0.01,0.05) & (0.10,0.14) & Skew & 1.027 & 1.000 & 0.647 & 0.741 & 0.947 & 0.979 & 0.890 & 0.928 \\ \hline
        ~ & ~ & Flat & 1.021 & 0.992 & 0.551 & 0.606 & 0.973 & 0.979 & 0.893 & 0.921 \\ \hline
        ~ & ~ & Bell & 1.033 & 0.998 & 0.834 & 0.959 & 0.964 & 0.845 & 0.894 & 0.889 \\ \hline
        (0.04,0.06) & (0.20,0.30) & Skew & 1.024 & 0.855 & 0.171 & 0.238 & 0.979 & 0.692 & 0.430 & 0.446 \\ \hline
        ~ & ~ & Flat & 1.022 & 0.868 & 0.252 & 0.336 & 0.981 & 0.800 & 0.698 & 0.701 \\ \hline
        ~ & ~ & Bell & 1.022 & 0.770 & 0.130 & 0.167 & 0.980 & 0.659 & 0.302 & 0.318 \\ \hline
        (0.15,0.30) & (0.04,0.06) & Skew & 1.029 & 1.059 & 0.434 & 0.495 & 0.998 & 0.945 & 1.069 & 1.071 \\ \hline
        ~ & ~ & Flat & 1.031 & 1.047 & 0.471 & 0.494 & 0.997 & 0.971 & 1.038 & 1.039 \\ \hline
        ~ & ~ & Bell & 1.036 & 1.009 & 0.394 & 0.487 & 0.998 & 0.871 & 1.190 & 1.204 \\ \hline
        (0.05,0.06) & (0.20,0.70) & Skew & 1.010 & 0.610 & 0.775 & 0.710 & 0.983 & 0.710 & 0.926 & 0.779 \\ \hline
        ~ & ~ & Flat & 0.994 & 0.517 & 0.905 & 0.551 & 0.991 & 0.794 & 0.982 & 0.805 \\ \hline
        ~ & ~ & Bell & 0.993 & 0.478 & 0.567 & 0.535 & 0.985 & 0.739 & 0.760 & 0.746 \\ \hline
    \end{tabular}}
\end{table}

\begin{table}[H]
\tbl{MSE ratios for Beta-Binomial success proportions $\bm{\alpha}=(\alpha_1,\ldots,\alpha_{10})$ and $\bm{\beta}=(\beta_1,\ldots,\beta_{10})$ comparing penalized parameter estimates to maximum likelihood for different penalization approaches.}
{\begin{tabular}{|c|c|c|c|c|c|c|c|c|c|c|}
    \cline{4-11}
        \multicolumn{3}{c|}{} & \multicolumn{4}{c|}{$\alpha_i$} & \multicolumn{4}{c|}{$\beta_i$}  \\ \hline
        $p_i \in (a_1,b_1)$ & $\nu_i \in (a_2,b_2)$ & Shape & Zero & Mean & Full & Min CV & Zero & Mean & Full & Min CV \\ \hline
        (0.05,0.10) & (4,6) & Skew & 0.993 & 0.922 & 0.201 & 0.319 & 1.037 & 0.718 & 0.097 & 0.192 \\ \hline
        ~ & ~ & Flat & 0.991 & 0.932 & 0.318 & 0.41 & 1.037 & 0.792 & 0.171 & 0.254 \\ \hline
        ~ & ~ & Bell & 0.992 & 0.922 & 0.144 & 0.209 & 1.039 & 0.662 & 0.071 & 0.119 \\ \hline
        (0.12,0.22) & (2,5) & Skew & 0.995 & 0.98 & 0.69 & 0.85 & 1.006 & 0.954 & 0.637 & 0.81 \\ \hline
        ~ & ~ & Flat & 0.997 & 1.015 & 0.999 & 1.015 & 1.006 & 1.01 & 0.999 & 1.01 \\ \hline
        ~ & ~ & Bell & 0.993 & 0.917 & 0.186 & 0.373 & 1.014 & 0.845 & 0.154 & 0.322 \\ \hline
        (0.17,0.22) & (3,8) & Skew & 0.992 & 0.928 & 0.593 & 0.746 & 1.017 & 0.832 & 0.544 & 0.664 \\ \hline
        ~ & ~ & Flat & 0.994 & 0.96 & 1.077 & 1.019 & 1.011 & 0.91 & 1.058 & 0.958 \\ \hline
        ~ & ~ & Bell & 0.991 & 0.931 & 0.283 & 0.401 & 1.021 & 0.791 & 0.241 & 0.333 \\ \hline
        (0.05,0.06) & (2,10) & Skew & 0.993 & 0.988 & 0.934 & 0.99 & 1.017 & 0.866 & 0.901 & 0.866 \\ \hline
        ~ & ~ & Flat & 0.996 & 0.958 & 1.015 & 0.96 & 1.007 & 0.787 & 1.007 & 0.788 \\ \hline
        ~ & ~ & Bell & 0.992 & 0.91 & 0.376 & 0.517 & 1.046 & 0.557 & 0.229 & 0.306 \\ \hline
    \end{tabular}}
\end{table}
\end{document}